\documentclass[twocolumn,a4paper]{IEEEtran} 

\newlength{\subfigwidth}
\usepackage[utf8]{inputenc}
\usepackage[english]{babel}
\usepackage{lipsum}
\usepackage{amssymb,amsmath}
\usepackage{booktabs}
\usepackage{tabularx}
\usepackage{subcaption}
\usepackage{graphicx}
\usepackage{adjustbox}
\usepackage{enumitem}
\usepackage{listings}
\usepackage{ textcomp }
\usepackage{float}
\usepackage[textsize=scriptsize]{todonotes}
\usepackage{color}
\usepackage[hidelinks]{hyperref}
\usepackage[numbers]{natbib}
\title{Automatic Library Version Identification, an Exploration of Techniques}

\author{
Thomas Rinsma \\
\url{thomasrinsma@gmail.com}\\
~\\
  \begin{tabularx}{.8\textwidth}{@{\extracolsep{\fill}}cc}
  Faculty of Science & Riscure \\
  Radboud University & Delft \\
  Nijmegen           & The Netherlands \\
  The Netherlands    &
  \end{tabularx}
}

\date{}

\begin{document}
\maketitle

\begin{abstract}
This paper is the result of a two month research internship on the topic of library version identification. In this paper, ideas and techniques from literature in the area of binary comparison and fingerprinting are outlined and applied to the problem of (version) identification of shared libraries and of libraries within statically linked binary executables. Six comparison techniques are chosen and implemented in an open-source tool which in turn makes use of the open-source \emph{radare2} framework for signature generation. The effectiveness of the techniques is empirically analyzed by comparing both artificial and real sample files against a reference dataset of multiple versions of dozens of libraries. The results show that out of these techniques, readable string--based techniques perform the best and that one of these techniques correctly identifies multiple libraries contained in a stripped statically linked executable file.
\end{abstract}

\section{Introduction}
When scrutinizing the security of an embedded device, one of the things an analyst looks for are vulnerabilities in the collection of executables and shared libraries contained in the device's firmware image. Often, these shared libraries are common, non-proprietary utility libraries and the executables will have often been statically linked against such libraries. A vulnerability in one of these libraries could lead to exploitation of the device, especially when considering that these libraries often process user controlled data (e.g. for multimedia decoding or compression). Some of these libraries have a history of vulnerable versions for which there are known exploits. The task of exploitation could therefore be as simple as determining the exact version or variant of each library and -- if a vulnerable version is found -- using public vulnerability information to devise an exploit for an executable file that makes use of the vulnerable library.

This paper focuses on a specific aspect of this process: the problem of automatically identifying the version of a given shared library file. Six techniques of three different signature types have are implemented in an experimental open-source tool in order to evaluate their effectiveness. Preliminary experiments are also performed on the applicability of these techniques on the harder problem of identifying libraries in statically linked executables. Samples and corresponding reference libraries of both ARM and MIPS architectures are used for these experiments.

\section{Related work}
\label{sec:relwork}

Intuitively, the identification problem as described above -- at least the shared library identification variant -- comes down to the problem of determining \emph{executable object} similarity. After all, in order to identify the best matching library version to some sample file, one has to calculate some metric of similarity between the sample file and each version of each reference library. In the case of statically linked executables, this will not result in matches of (close to) 100\% similarity, but it will give an indication of the amount of overlap between a library version and the sample file, which might be an indicator for the probability that that library (version) is actually contained inside the sample file.

A structural way to perform such an executable object comparison was laid out by Dullien and Rolles in \cite{dullien2005graph}. They reduce it to the problem of creating a (fuzzy) graph isomorphism between control-flow graphs (CFGs) of both executable objects. Specifically, they treat an executable object as a \emph{graph of graphs}, i.e.~a function \emph{callgraph} where each of the nodes -- i.e.~each function -- contains that function's control-flow graph which in turn has basic-blocks as its nodes. Metrics are defined to determine basic-block and function similarity, allowing a full isomorphism to be created.

Building on these techniques and specifically focusing on polymorphic worm detection, the authors in \cite{Kruegel2006} apply a fingerprinting technique on a canonical representation of $k$-subgraphs of a CFG, allowing quick (partial) matching of a sample file. Following up on this, Cesare and Xiang \cite{Cesare2011} implement a complete malware classification system which among other things translates full CFGs into strings based on a specific grammar, allowing them to be compared quickly using Levenshtein distance (i.e.~edit distance). Similarly, Koret's open source malware clustering toolkit \emph{Cosa Nostra} represents CFGs as prime products determined by the cyclomatic complexity values of functions, allowing for permutation-independent fuzzy comparisons \cite{Koret2016}.

Doing away with graph-theoretic techniques, Gheorghescu proposes three classification methods on the flat list of basic-blocks of an executable object, the most innovative of which uses Bloom filters (see section \ref{sec:tech-bloom}) to achieve a fixed small signature size \cite{Gheorghescu2005}.

Taking an entirely different approach, Tian et al.\ propose a system for malware classification that achieves high accuracy by using the printable strings from a sample as an input for several classification algorithms including Bayesian and K-nearest neighbour \cite{Tian2009}.

The work presented in this paper is a preliminary exploration of the effectiveness of applying variants of the above-mentioned techniques to the problem of library version identification.

\section{CFG generation tools}
\label{cfg_tools}
%

To perform the non-string based comparison techniques mentioned in section \ref{sec:relwork} (i.e.~all but the system by Tian et al.), we first need to construct the CFGs of all of the functions in the executable objects in question. This requires disassembling the objects and using knowledge of the instruction set and calling conventions in order to build a directed graph of the control flow between basic-blocks. Several pieces of software can perform this task, the most popular of which is the proprietary \emph{IDA Pro}, which is easily scriptable and is often used for similar purposes. An interesting alternative for our purpose however is the open source binary analysis framework \emph{angr} \cite{Koret2016}. Like IDA Pro, it supports many architectures and formats and in addition to its ability to generate accurate CFGs using symbolic execution, it has built-in binary comparison functionality based on \cite{dullien2005graph} (see section~\ref{sec:tech-graph-isomorphism}).
Another alternative is the \emph{radare2} reverse engineering framework. It too has built-in support for (static) CFG generation and binary comparison (see section~\ref{sec:tech-graph-isomorphism}). Additionally, it supports even more architectures than \emph{angr} and has bindings for almost any language. In practice, it also turns out to be faster than \emph{angr} in static analysis tasks. For this reason \emph{radare2} was used in lieu of \emph{angr} for the final iteration of this project  as is explained in more detail in section \ref{sec:tech-graph-isomorphism}.


\section{Techniques}
\label{sec:tech}
From the related work discussed in section \ref{sec:relwork} we can distill multiple techniques of creating executable object \emph{signatures} (i.e. \emph{fingerprints}) and multiple techniques of comparing those signatures. Three of these combinations are detailed in this section: basic block hash comparison using Bloom filters (section~\ref{sec:tech-bloom}), several \emph{cyclomatic complexity}--based techniques (section~\ref{sec:tech-cc}) and several string-based techniques (section~\ref{sec:tech-strings}). Before discussing these techniques, we must however understand why they are useful, especially when compared with graph isomorphism-based techniques. This is explained in section~\ref{sec:tech-graph-isomorphism}.

\subsection{Graph isomorphisms}
\label{sec:tech-graph-isomorphism}
While \emph{angr} has the ability to perform several static analysis techniques, its main purpose is symbolic execution. For generating the CFG of an executable object, it offers two methods: \texttt{CFGAccurate} and \texttt{CFGFast}. The former uses symbolic execution of basic-blocks to accurately determine all of the possible control-flow paths, while the latter uses a more traditional heuristics-based approach. The result of either can be passed to \emph{angr}'s \texttt{BinDiff} analysis method which is named after commercial software of the same name by Zynamics \cite{BinDiffZynamics}. Just like its namesake, it implements the graph isomorphism technique from \cite{dullien2005graph}.

For our purposes, we can use these methods to create a graph isomorphism between two potentially similar libraries. A similarity measure can then be derived from the number of identical or \emph{almost identical} functions and basic-blocks.

However, such an isomorphism has to be created for every reference file (i.e.\ every library version to compare against). Ideally we would want to compare a sample against \emph{every version of every library} in our reference set, and possibly even against compilations of the same version by different compilers, or with varying optimization levels or other compiler flags\footnote{Optimistically, such compiler and flag differences would be abstracted away by techniques such as \cite{dullien2005graph} but as those authors themselves note, modern optimizing compilers can make drastic changes to assembly output depending on compiler flags and other variables.}. This means that the number of isomorphism calculations that needs to be done for one sample is really high and will grow quickly (but linearly) with added reference libraries. Additionally, the CFG information (i.e.\ the whole graph and associated data) needed by \emph{angr}'s \texttt{BinDiff} needs to be pre-calculated and stored for each one of these reference files and loaded into memory when the comparison is performed.

As it turns out, both of these factors make \emph{angr} a less than ideal framework for our purpose. The data structure produced by its CFG generation methods is large and contains data unneeded for the \texttt{BinDiff} method. Additionally, the \texttt{BinDiff} method makes heavy use of \emph{angr}'s \emph{lifting} functionality to lift machine code to an intermediate representation (using PyVEX), which takes up most of the method's long execution time. An attempt was made to mitigate these problems by stripping down both the CFG datastructure and simplifying \texttt{BinDiff} by performing the lifting at CFG generation time. This resulted in a small improvement but the core problems of heavy memory usage and long computation times remain. Both of these are however not necessarily a problem of \emph{angr} but of the graph isomorphism-based algorithm in general. The \texttt{radiff2} tool of \emph{radare2} also implements the algorithm from Dullien and Rolles \cite{dullien2005graph} and achieves similar performance. It becomes clear why \cite{Cesare2011} and \cite{Kruegel2006} implement more efficient CFG fingerprinting techniques: graph isomorphisms are generally too slow for a one-to-many comparison. This is especially true for \emph{angr}'s implementation which is slowed down by operations like the \emph{lifting} described above. This is not surprising however, because \emph{angr} is mainly intended to be used interactively on very small (and often artificial) samples.

For these reasons, both the original and the optimized version of \emph{angr}'s \texttt{BinDiff} were not integrated into the library identification tool. They are not practically usable with a reference database of a few hundred samples\footnote{A single comparison with \emph{angr}'s \texttt{BinDiff} can take up to a minute for large reference libraries and a lot longer for even larger samples.} and will therefore not scale to much larger databases that could be used in practice.

\subsection{Basic-block matching using Bloom filters}
\label{sec:tech-bloom}
Clearly we need signatures that are both quicker to compare against and have a smaller footprint than a full CFG. Gheorghescu's Bloom filter approach \cite{Gheorghescu2005} satisfies these requirements. The technique consists of creating a fixed size Bloom filter\footnote{Bloom filter: a fixed size bit-array populated by ones at positions corresponding to the hashes of its containing elements. It is normally used as a data structure for fast probablistic element look-up.} \cite{bloom1970space} for each reference object and inserting into it the hashes of all of that file's basic-blocks. It is not specified exactly what the input to the filter's hash function is, but presumably the raw byte code of a basic-block is used. We can now obtain a similarity ratio between files by calculating the \emph{Jaccard index} (or \emph{similarity coefficient}) of the sets of bits in their Bloom filters, i.e. for filters $x$ and $y$, calculate $d(x,y) = \Sigma_i (x_i \wedge y_i) / \Sigma_i (x_i \vee y_i)$. Because the run time of element insertion and look-up in a Bloom filter are both $O(1)$, this makes the time complexity of signature generation linear in file size. More importantly, signature comparison is constant in the size of the Bloom filter.

We can implement this technique by using \emph{radare2} to obtain basic-block information. Specifically, we can use the \emph{type} label it gives to every instruction in place of the instruction byte code itself. This way, we abstract away differences in register allocation and constants. The hash of basic-block $B$ consisting of instructions $\{b_1,...,b_n\}$ is then constructed as follows:
$$ h(B) = \text{crc32}(\text{\emph{type}}(b_0)\;||\;\text{\emph{type}}(b_1)\;||\;...\;||\;\text{\emph{type}}(b_n)) $$
For a Bloom filter of size $m$ bits, the $log_2(m)$ least significant bits of the hash output are used as an index. A value of $m=2^{15}$ is used in this implementation.

\subsection{Function matching using cyclomatic complexity}
\label{sec:tech-cc}
Another way to create really small signatures (i.e. in the order of $O(\log n)$ in the file-size) of executable objects is by calculating and comparing the cyclomatic complexity values of their containing functions. The cyclomatic complexity $M_f$ of a function $f$ whose CFG contains $N_f$ nodes (i.e. basic-blocks) and $E_f$ edges is usually defined as: $M_f = E_f - N_f + 2$. In the implementation of this technique by Koret in the \emph{Cosa Nostra} framework, the signature of an executable object is then calculated by taking the product of primes, indexed by the cyclomatic complexity of each function \cite{Koret2016}. In other words, if $p_n$ denotes the $n$th prime, the file signature is calculated by:
$$ \text{sig}(\text{file}) = \prod_{f \in \text{file}} p_{\scriptscriptstyle M_f}$$

Two of these signatures or \emph{fuzzy hashes} can then be compared by factoring them and counting the number of differing factors between them. This concept of using \emph{small primes products} to determine if one feature-set is (almost) a permutation of another was introduced by Dullien and Rolles \cite{dullien2005graph}. In their algorithm it is instead used on sets of instructions for determining code similarity as a step in finding a graph isomorphism.

Retrieving the needed information to generate such signatures is trivial when using \emph{radare2}. After analysing a file it provides the cyclomatic complexity value as the \texttt{cc} attribute of every function in the output of the \texttt{aflj} command. Storing not only the prime product but also the list of \texttt{cc} values (or the list of prime factors, as used for technique \texttt{cc3} in our tool) for every reference file saves a factoring step and allows us to perform additional types of comparison on them. For this project, two additional comparison techniques were implemented on the list of \texttt{cc} values: Levenshtein distance (permutation-sensitive) as technique \texttt{cc1} and set similarity (permutation-insensitive) as technique \texttt{cc2}.

\subsection{Printable strings matching}
\label{sec:tech-strings}
Every non-trivial executable object contains a set of printable strings. These sets often consist of error messages, copyright or usage information, but also of strings for internal use like file content in media parsing or generation libraries. Additionally, symbol tables can be present, containing function or object names. Tian et al.\ show that such a list of strings can be an accurate signature of an executable object when used for malware classification using various machine learning algorithms \cite{Tian2009}. However, one wonders if it is possible to forgo such algorithms and instead perform a fuzzy comparison using Levenshtein distance, or even more trivial, a simple set difference calculation.

Retrieving the printable strings from an executable object is a trivial operation. The basic functionality of the Unix \texttt{strings} command can be replicated in less than 50 lines of code. As a trade-off, the resulting signature (i.e.~the list of strings) is not as small as those in the previous two techniques, nor is it constant in size. It is however never bigger than the executable object itself (which can be the case when using \emph{angr}'s \texttt{BinDiff} comparison method). On these signatures, two comparison techniques are implemented and tested:
\begin{itemize}[noitemsep,topsep=0pt,parsep=0pt,partopsep=0pt]
  \item \textbf{Fuzzy} string matching by calculating the Levenshtein distance on two sorted and concatenated lists of strings. In other words, calculate $a_{cat}$ and $b_{cat}$ as the concatenations of all of the strings from the sample and the reference file respectively, before calculating the similarity ratio as follows: $$r = 1 - \frac{\text{Levenshtein}(a_{cat}, b_{cat})}{\text{max}(|a_{cat}|, |b_{cat}|)}$$
  This metric handles both differences within strings and differences between the lists of strings, i.e. added or removed strings. In the tool, this is implemented as technique \texttt{str1}.

  \item \textbf{Exact} position-sensitive string matching using the similarity between two sets of strings. Specifically, the similarity ratio between two string sets $A$ and $B$ is calculated as:
  $$r = \frac{|A \cup B|}{|A \cap B|} $$ 
  This will give the \emph{Jaccard index} of the sets, indicating the ratio of strings that occur in both the sample and the reference file versus all of the strings that occur in either file. This is implemented as technique \texttt{str2}.
\end{itemize}

\section{The library identification tool}
The aforementioned techniques (sections~\ref{sec:tech-bloom} - \ref{sec:tech-strings}) have been implemented into an experimental open-source \emph{library identification} tool\footnote{\url{https://github.com/Riscure/Library-Identification/}}.
It is written in Python, making use of several open-source packages including the \texttt{r2pipe} package for communicating with a \emph{radare2} instance during signature generation. Architecturally, the tool is composed of a main file, \texttt{library\_identification.py}, which contains the \texttt{LibraryFile} and \texttt{ReferenceDB} classes which respectively implement an abstraction for shared library files, and a manager for library signature databases. These classes are used by the tool's two front-end scripts: \texttt{identify.py} and \texttt{generate\_db.py}.

Six signature comparison techniques are implemented in \texttt{identify.py} as functions with a standardised interface, allowing them to be passed as function pointers to helper functions. Table \ref{table:comp_tech} shows all of these comparison functions with descriptions of what they do.

\begin{table*}[t]
\begin{adjustbox}{center}
\small
\begin{tabular}{@{}p{6.5cm}p{.8cm}p{8cm}@{}}
\toprule
\multicolumn{1}{c}{Function name} & \multicolumn{1}{c}{Short name} & \multicolumn{1}{c}{Description} \\ \midrule

\texttt{compare\_bb\_hash\_bloomfilter} & \texttt{bloom} & Compares the sample and the reference by comparing the Bloom filter signatures of each using the Jaccard index. \\ 
\texttt{compare\_cc\_list\_levenshtein} & \texttt{cc1} & Compares the cyclomatic complexity values of all functions in the sample with those of all functions in the reference by taking the Levenshtein distance between these lists. \\
\texttt{compare\_cc\_list\_set\_union} & \texttt{cc2} & Performs a permutation-independent comparison between the lists of cyclomatic complexity values of both the sample and the reference by treating both as sets and calculating the overlap between these sets using the Jaccard index. \\
\texttt{compare\_cc\_spp} & \texttt{cc3} & Compares the cyclomatic complexity values of the functions in the sample and the reference by factoring the small prime products of each and determining the number of matching factors. \\
\texttt{compare\_strings\_concat\_levenshtein} & \texttt{str1} & Performs a fuzzy string and string list comparison by taking the Levenshtein distance between the concatenations of the lists of strings in both the sample and the reference. \\
\texttt{compare\_strings\_set\_union} & \texttt{str2} & Performs an exact, permutation-independent string list comparison by treating the lists of strings in both the sample and the reference as sets and calculating the overlap between these sets using the Jaccard index. \\\bottomrule
\end{tabular}
\end{adjustbox}
\caption{Implemented comparison techniques}
\vspace{.5cm}
\label{table:comp_tech}
\end{table*}

\section{Reference dataset}

The implemented techniques have been tested on both types of samples: shared library files and statically linked executables. For the latter category, both real and semi-artificial samples were picked. Information about these samples and the results obtained from them can be found in section~\ref{sec:results}.

To provide the best testing scenario, a database of many library version signatures is needed. For the experiments in this paper, a database was constructed using a custom tool that downloads and cross-compiles as many versions as possible of a given open-source library. This was done by making use of the build instructions in \texttt{PKGBUILD} files provided for each package in the Arch Linux repositories\footnote{This approach was preferred over a binary source (e.g.\ Debian package archives) because it allows us to create binaries of old library versions without future security patches being applied.}.
The result after applying this tool to a selection of common libraries and other packages starting with \texttt{lib} is a set of more than 60 libraries for ARM and MIPS, each with anywhere between 1 and 40 versions, but with an average of about 5. The full dataset used in the experiments is provided in appendix A.


\section{Results}
\label{sec:results}
In the sections below, the results of several experiments are shown. Section \ref{sec:results-shared} details the results of several experiments related to the identification of shared libraries, while section~\ref{sec:results-static} deals with the secondary goal of version identification of libraries contained within statically linked binaries.
\subsection{Identifying shared libraries}
This section contains explanations and the results of several experiments with shared libraries as the sample files.
\label{sec:results-shared}
\subsubsection{Speed}
\label{sec:results-speed}
After running the tool on various samples, the first thing that stands out is the major difference in speed between the techniques. Table \ref{table:duration-mips} shows the duration of using each technique to compare a sample to 188 references (the total amount of library versions in the dataset for the MIPS architecture). We can see that \texttt{str1} --- the fuzzy string matching technique --- performs the slowest, while \texttt{cc1} and \texttt{cc2} perform the fastest. In fact, for all of the samples tested, fuzzy string matching is slower than \texttt{cc1} and \texttt{cc2} by a factor of more than a thousand. The third cyclomatic complexity-based technique --- \texttt{cc3} --- is much slower than the other two. This is to be expected because it employs a prime factorisation algorithm which is relatively computationally expensive. Lastly, we observe that the Bloom filter comparison algorithm takes more time than than \texttt{cc1} and \texttt{cc2}, but still finishes in well under a second on our database of 188 references. More importantly, its running time varies very little between the different samples and does not depend on their size.

\begin{table*}[h]
\begin{adjustbox}{center}
\centering
\small
\begin{tabular}{@{}cc|rrrrrr@{}}
\toprule
\multicolumn{2}{c|}{Sample library}
              & \multicolumn{1}{c}{\texttt{str1}}
              & \multicolumn{1}{c}{\texttt{str2}}
              & \multicolumn{1}{c}{\texttt{cc1}}
              & \multicolumn{1}{c}{\texttt{cc2}}
              & \multicolumn{1}{c}{\texttt{cc3}}
              & \multicolumn{1}{c}{\texttt{bloom}}  \\ \midrule
libpng 1.6.15 & 721 KB &  91903 ms & 47 ms &  9 ms & 15 ms & 13095 ms & 304 ms  \\
libz 1.1.3    & 208 KB &  16257 ms & 28 ms &  5 ms &  9 ms & 13210 ms & 364 ms  \\
curl 7.43.0   & 380 KB & 146191 ms & 80 ms & 66 ms & 19 ms & 13437 ms & 309 ms  \\
bzip2 1.0.6-5 & 157 KB &  18625 ms & 28 ms &  5 ms & 10 ms & 13152 ms & 348 ms  \\
\bottomrule
\end{tabular}
\end{adjustbox}
\caption{Averaged duration of the comparison of each sample against 188 MIPS library versions on a Core i7 laptop.}
\vspace{.5cm}
\label{table:duration-mips}
\end{table*}

\subsubsection{Version identification}
\label{sec:results-version-ident}
When the identification tool is applied to a sample library version of which the signature is present in the reference database, all six techniques produce the same version as a match with 100\% similarity, as would be expected. More interestingly, when comparing a sample library version to other versions of the same library, we see that with most techniques, the versions with the highest similarity scores are closest in version to the sample. An example of this is shown in table \ref{table:libjpeg-similarities}. Here, libjpeg version 9.2.0 is compared to all libjpeg versions in the database. We see that the nearest version --- 9.1.0 --- is given the second-highest similarity score by all the techniques. In fact, for all techniques except for \texttt{cc2}, the sorted list of versions by similarity score is in perfect chronological order. Assuming that each version is based on the previous version and no major rewrites were performed, this result is entirely as expected. The fact that in this case, all techniques give a similarity score of 93+\% to the version nearest to the sample, means that even if the exact version of the sample is not in the reference database, a good approximation of its version will still be returned.

\begin{table*}[h]
\begin{adjustbox}{center}
\centering
\small
\begin{tabular}{@{}c|rrrrrr@{}}
\toprule
\multicolumn{1}{c|}{Reference library}
              & \multicolumn{1}{c}{\texttt{str1}}
              & \multicolumn{1}{c}{\texttt{str2}}
              & \multicolumn{1}{c}{\texttt{cc1}}
              & \multicolumn{1}{c}{\texttt{cc2}}
              & \multicolumn{1}{c}{\texttt{cc3}}
              & \multicolumn{1}{c}{\texttt{bloom}}  \\ \midrule
libjpeg 9.2.0 &  100.00\% & 100.00\% & 100.00\% & 100.00\% & 100.00\% & 100.00\% \\
libjpeg 9.1.0 &  99.97\%  & 99.35\%  & 98.75\%  & 93.62\%  & 97.49\%  & 97.19\%  \\
libjpeg 9.0.0 &  99.87\%  & 94.62\%  & 91.41\%  & 84.00\%  & 90.48\%  & 88.38\%  \\
libjpeg 8.4.0 &  99.39\%  & 92.91\%  & 89.66\%  & 70.37\%  & 88.09\%  & 84.64\%  \\
libjpeg 8.3.0 &  99.25\%  & 92.77\%  & 86.83\%  & 70.37\%  & 86.52\%  & 78.78\%  \\
libjpeg 8.0.2 &  99.24\%  & 92.77\%  & 86.83\%  & 73.58\%  & 86.52\%  & 78.42\%  \\
libjpeg 8.0.1 &  99.22\%  & 92.77\%  & 86.21\%  & 73.58\%  & 86.52\%  & 78.06\%  \\
libjpeg 8.0.0 &  99.20\%  & 92.61\%  & 86.21\%  & 73.58\%  & 86.52\%  & 77.34\%  \\
libjpeg 7.0.0 &  95.14\%  & 88.85\%  & 78.37\%  & 65.45\%  & 78.37\%  & 77.12\%  \\
\bottomrule
\end{tabular}
\end{adjustbox}
\caption{Similarity of \texttt{libjpeg.so.9.2.0} (MIPS) to other libjpeg versions.}
\vspace{.5cm}
\label{table:libjpeg-similarities}
\end{table*}

\subsubsection{False positives}
\label{sec:results-false-pos}
While a good comparison technique for our purpose should give a high similarity score to close versions of the same library, this is not the only metric of effectiveness. In table \ref{table:similarity-second-highest-match-mips}, another metric is shown: \emph{the similarity score of the highest rating library version that is not a version of the same library as the sample}. In other words, all results of the correct library are ignored, and the highest remaining similarity score is recorded. Since these results are by definition not matches or near-matches\footnote{This is assumed to be true in this dataset. In general, libraries could share code or be forked from each other, in which case high similarity ratings could be legitimate.}, the recorded scores should be low. In practice, \texttt{str2} performs really well in this regard: for all samples in table \ref{table:similarity-second-highest-match-mips}, the highest ``false positive'' similarity score is below 9\%, which is low compared to the 20+\% values returned by other techniques. The cyclomatic complexity--based techniques perform badly because the probability that functions from unrelated libraries have the same cyclomatic complexity value is high, especially for large libraries. This is an expected trade-off of the small signature size obtained with these techniques. The \texttt{bloom} technique performs similarly, largely because of a similar reason: unrelated libraries share basic-blocks that produce the same hash. Considering the fact that the hash is made from the list of instruction \emph{types}, and that these references contain many small basic-blocks, this was to be expected.

An outlier here is the result of \texttt{bloom} on libz version 1.1.3. In this case, all non-libz library versions are given similarity scores of less than 1\%, a \emph{much} better result than the others in its column. Why only this specific library shares so little basic-block hashes with others is unknown. One possible cause of this could be that instructions in versions of this library are consistently different or ordered differently from those in other libraries, which could be explained by a different compiler or different compiler flags being used.

\begin{table*}[h]
\begin{adjustbox}{center}
\centering
\small
\begin{tabular}{@{}c|rrrrrr@{}}
\toprule
\multicolumn{1}{c|}{Sample library}
              & \multicolumn{1}{c}{\texttt{str1}}
              & \multicolumn{1}{c}{\texttt{str2}}
              & \multicolumn{1}{c}{\texttt{cc1}}
              & \multicolumn{1}{c}{\texttt{cc2}}
              & \multicolumn{1}{c}{\texttt{cc3}}
              & \multicolumn{1}{c}{\texttt{bloom}}  \\ \midrule
libpng 1.6.15 & 23.16\% & \textbf{3.96\%}  & 30.00\%  & 54.29\%  & 64.15\%  & 27.78\%  \\
libz 1.1.3    & 20.19\% & \textbf{7.71\%}  & 29.01\%  & 69.57\%  & 46.15\%  & 0.97\%   \\
curl 7.43.0   & 22.28\% & \textbf{2.89\%}  & 37.30\%  & 62.12\%  & 73.40\%  & 38.89\%  \\
bzip2 1.0.6-5 & 20.17\% & \textbf{8.27\%}  & 29.63\%  & 57.58\%  & 33.93\%  & 31.82\%  \\
\bottomrule
\end{tabular}
\end{adjustbox}
\caption{Similarity of the highest matching version of an incorrect library.}
\label{table:similarity-second-highest-match-mips}
\end{table*}

\subsection{Identifying libraries within statically linked executables}
\label{sec:results-static}
All six techniques perform well at the task of identifying the most likely version of a shared library file. We would, however, also like to know how well these techniques apply to the problem of identifying libraries inside statically linked executables. In order to evaluate this, the tool was run on both artificial and real-world statically linked executables. The results of which are detailed here.

Firstly, the sample code provided by libjpeg in \texttt{example.c} was used to create a dummy program that calls the \texttt{read\_JPEG\_file} and \texttt{write\_JPEG\_file} functions. Nine samples were created by statically linking this program with nine different versions of libjpeg. Each sample was stripped of symbols before performing the experiments. Table \ref{table:libjpeg-example-static-tests} shows for each combination of sample and technique, whether the correct libjpeg version (the one that the sample was linked with) is the top result (i.e. whether it has the highest similarity score out of all the references).

In the table, we can see that \texttt{str2} is the only technique that consistently gives the highest similarity score to the correct library version. The fuzzy string comparison technique, \texttt{str1}, performs almost as well but strangely produces version 7.0.0 for the sample which was linked to libjpeg version 8.0.0. 

Of course, when interpreting the program output in this manner, we assume that the sample file is statically linked to exactly one library and that this library is contained in the reference database. A more complex scenario would be a sample that is statically linked against multiple libraries. One such a sample is the statically linked version of the \texttt{castget} program, an open source podcast downloader\footnote{From \url{http://castget.johndal.com/}}. Three libraries are statically linked into this relatively large (1.7MB) binary: curl (7.50.0), libxml2 (2.8.0) and id3lib (3.8.3). Because two out of three of these libraries are in the reference database -- curl and id3lib -- it makes for a good real-world test case.

Looking at the results, we see that \texttt{str2} is the only technique where the two correct matches have the highest similarity scores, with a significant percentage drop afterwards (i.e. from 11.13\% for id3lib to 2.51\% for libmp4v2). The \texttt{str2} technique has matched the correct version of curl but not of id3lib, where it matched version 3.8.0 instead of 3.8.3, the actual version contained in the sample. This was however entirely expected, because version 3.8.0 is the only reference version for id3lib in the database.

\newcommand{\no}{no}
\newcommand{\yes}{\textbf{yes}}
\newcommand{\wrongver}{}
\begin{table*}[t]
\begin{adjustbox}{center}
\centering
\small
\begin{tabular}{@{}c|cccccc@{}}
\toprule
\multicolumn{1}{c|}{Sample file}
              & \multicolumn{1}{c}{\texttt{str1}}
              & \multicolumn{1}{c}{\texttt{str2}}
              & \multicolumn{1}{c}{\texttt{cc1}}
              & \multicolumn{1}{c}{\texttt{cc2}}
              & \multicolumn{1}{c}{\texttt{cc3}}
              & \multicolumn{1}{c}{\texttt{bloom}}  \\ \midrule
\texttt{jpeg\_7.0.0\_example\_static} & \yes & \yes & \yes & \yes & \yes & \yes  \\
\texttt{jpeg\_8.0.0\_example\_static} & \wrongver (7.0.0) & \yes & \wrongver (8.0.1) & \no  & \wrongver (multiple) & \wrongver (multiple)  \\
\texttt{jpeg\_8.0.1\_example\_static} & \yes & \yes & \yes & \no & \yes & \yes  \\
\texttt{jpeg\_8.0.2\_example\_static} & \yes & \yes & \wrongver (multiple) & \no & \wrongver (multiple) & \wrongver (multiple)  \\
\texttt{jpeg\_8.3.0\_example\_static} & \yes & \yes & \yes & \no & \wrongver (8.0.1) & \wrongver (8.0.1) \\
\texttt{jpeg\_8.4.0\_example\_static} & \yes & \yes & \wrongver (8.3.0) & \no & \wrongver (multiple) & \wrongver (multiple)  \\
\texttt{jpeg\_9.0.0\_example\_static} & \yes & \yes & \wrongver (8.3.0) & \yes & \wrongver (multiple) & \wrongver (7.0.0)  \\
\texttt{jpeg\_9.1.0\_example\_static} & \yes & \yes & \wrongver (multiple) & \yes & \yes & \yes  \\
\texttt{jpeg\_9.2.0\_example\_static} & \yes & \yes & \wrongver (multiple) & \wrongver (multiple) & \yes & \yes  \\

\bottomrule
\end{tabular}
\end{adjustbox}
\caption{Does the correct libjpeg version have the highest similarity rating when comparing a statically linked executable against 188 references? ``\textbf{yes}'': the correct libjpeg version has the highest similarity score; ``\textbf{no}'': a different library has the highest similarity score; ``\textbf{(\emph{n})}'': version \emph{n} of libjpeg has the highest similarity score; ``\textbf{(multiple)}'': several versions of libjpeg have a shared highest similarity score.}
\vspace{.5cm}
\label{table:libjpeg-example-static-tests}
\end{table*}

\begin{table*}[t]
  \small
    \begin{subtable}{\subfigwidth}
      \centering
        \begin{tabular}[c]{ll|r}
        \toprule
        \multicolumn{1}{c}{Library} & \multicolumn{1}{c|}{Version} & \multicolumn{1}{c}{Similarity} \\
        \midrule
        libisoburn &  1.4.0   & 18.03\% \\
        \textbf{curl}       &  \textbf{7.50.0}  & 17.85\% \\
        libisofs   &  1.4.6   & 16.95\% \\
        libarchive &  3.2.2   & 16.53\% \\
        libexif    &  0.6.21  & 14.93\% \\
        \bottomrule
        \end{tabular}
        \caption{\texttt{str1}}
    \end{subtable}
    \vspace{.5cm}
    \begin{subtable}{\subfigwidth}
      \centering
        \begin{tabular}[c]{ll|r}
        \toprule
        \multicolumn{1}{c}{Library} & \multicolumn{1}{c|}{Version} & \multicolumn{1}{c}{Similarity} \\
        \midrule
        \textbf{curl}     &  \textbf{7.50.0} & 18.91\% \\
        \textbf{id3lib}   &  3.8.0  & 11.13\% \\
        libmp4v2 &  2.0.0  &  2.51\% \\
        libebml  &  1.3.1  &  1.48\% \\
        libburn  &  1.2.2  &  1.37\% \\
        \bottomrule
        \end{tabular}
        \caption{\texttt{str2}}
    \end{subtable}
    \begin{subtable}{\subfigwidth}
      \centering
        \begin{tabular}[c]{ll|r}
        \toprule
        \multicolumn{1}{c}{Library} & \multicolumn{1}{c|}{Version} & \multicolumn{1}{c}{Similarity} \\
        \midrule
        libmp4v2   &  2.0.0      & 44.98\% \\
        libisofs   &  1.4.2      & 43.39\% \\
        libarchive &  3.2.1      & 30.36\% \\
        libisoburn &  1.3.6      & 30.19\% \\
        \textbf{curl} &  7.37.1  & 29.74\% \\
        \bottomrule
        \end{tabular}
        \caption{\texttt{cc1}}
    \end{subtable}
    \begin{subtable}[c]{\subfigwidth}
      \centering
        \begin{tabular}[c]{ll|r}
        \toprule
        \multicolumn{1}{c}{Library} & \multicolumn{1}{c|}{Version} & \multicolumn{1}{c}{Similarity} \\
        \midrule
        libisoburn &  1.4.6  & 56.07\% \\
        libarchive &  3.2.0  & 50.47\% \\
        opus       &  1.1.3  & 46.46\% \\
        libpng     &  1.5.13 & 45.92\% \\
        libmodplug &  0.8.7  & 45.87\% \\
        \bottomrule
        \end{tabular}
        \caption{\texttt{cc2}}
    \end{subtable}
    \begin{subtable}[c]{\subfigwidth}
      \centering
        \begin{tabular}[c]{ll|r}
        \toprule
        \multicolumn{1}{c}{Library} & \multicolumn{1}{c|}{Version} & \multicolumn{1}{c}{Similarity} \\
        \midrule
        libisofs   & 1.4.0  & 53.94\% \\
        libarchive & 3.2.0  & 42.94\% \\
        libisoburn & 1.4.6  & 42.94\% \\
        \textbf{curl} & 7.37.1 & 31.90\% \\
        libvisual  & 0.4.0  & 29.31\% \\
        \bottomrule
        \end{tabular}
        \caption{\texttt{cc3}}
    \end{subtable}
    \begin{subtable}[c]{\subfigwidth}
      \centering
        \begin{tabular}[c]{ll|r}
        \toprule
        \multicolumn{1}{c}{Library} & \multicolumn{1}{c|}{Version} & \multicolumn{1}{c}{Similarity} \\
        \midrule
        libdca        &  0.0.5  & 60.71\% \\
        libsamplerate &  0.1.3  & 51.72\% \\
        libmspack     &  0.0.2...\footnotemark  & 37.50\% \\
        libupnp       &  1.6.18 & 37.50\% \\
        bzip2         &  1.0.6  & 35.56\% \\        
        \bottomrule
        \end{tabular}
        \caption{\texttt{bloom}}
    \end{subtable}
    
    \caption{Top 5 distinct matched libraries for the \texttt{castget} ARM sample. Each table shows the result for a single technique.}
    \vspace{.5cm}
    \label{table:castget-results}
\end{table*}

\section{Robustness against obfuscation}

While some of the techniques are clearly more effective than others when applied to the samples above, it is also good to consider their robustness against intentional and unintentional obfuscation in more obscure scenarios. The authors of firmware images containing shared libraries or statically linked executables might exploit weaknesses of the identification techniques in order to make it as difficult as possible for an analyst to determine the version of these libraries. A clear example of such a \emph{countermeasure} is string obfuscation. There exists a wide range of techniques for string obfuscation -- ranging from trivial transformations to cryptographic techniques -- but even the simplest of these will render \texttt{str1} and \texttt{str2} fruitless. On the other hand, the cyclomatic complexity and basic-block--based techniques are sensitive to changes to the CFG structure or individual instructions which can be caused by (heavy) compile-time optimization or intentional code obfuscation. In such situations, the string-based techniques are the most robust.

While such countermeasures are important to keep in mind, the best-case scenario of a non-optimized and non-obfuscated sample was assumed during this project. Certainly, it is much more straight-forward for a firmware-creator to keep their included open-source libraries up to date than to attempt to obfuscate them to hide the fact that they are out of date. Therefore one would not expect to come across many intentionally obfuscated open-source libraries in practice.


Another factor that could cause the string-based techniques to perform less well is the fact that the number of strings in libraries is not necessarily related to their size, i.e.~there could be low-level libraries that barely contain any (unique) strings, causing low similarity scores and relatively high false-positive scores.

\section{Future work}

The work presented in this paper only scratches the surface of what could be possible in terms of library version identification using existing fingerprinting and binary comparison techniques. The six implemented techniques represent several combinations of fingerprint \emph{creation} and fingerprint \emph{comparison} techniques, but more combinations and variations are possible. Specifically, more advanced approaches like the use of machine learning algorithms \cite{Tian2009} could yield better results.

The biggest problem with the current approach is that all techniques compare the entire contents of the sample file to the entire contents of reference files, resulting in \emph{absolute} rather than \emph{relative} similarity scores. These are useful when the sample is a shared library file but less so for statically linked executables, especially when these contain multiple linked libraries, as seen in table \ref{table:castget-results}. In such cases, one might want to first identify which parts of the executable file are library code at all and possibly even apply fingerprinting techniques on a per-function basis\footnote{IDA's \emph{F.L.I.R.T.}\cite{IDAFLIRT} provides library function identification using small fingerprints, but it is rather fragile, and geared more towards manual reverse-engineering and small sets of libraries.}. Research is required to determine whether or not such solutions are feasible and more effective.

More generally, other approaches to this problem could be the use of symbolic execution (i.e.~comparing functions or basic-blocks by their constraints) or visualisation \cite{conti2008visual} using image processing techniques.

\section{Conclusions}

In this project, six comparison techniques were detailed and implemented, inspired by existing research in the areas of executable file comparison and fingerprinting. Out of these techniques, the exact set-based readable string comparison technique, \texttt{str2}, performed the best. In terms of effectiveness it (empirically) outperforms all others, most notably by correctly assigning relatively high similarity scores to libraries contained inside a statically linked executable, and low scores to others. Using the output of this technique -- given an exhaustive reference database -- an analyst is able to more quickly determine (the versions of) libraries contained in statically linked executables.

While \texttt{str2} performs well in the the tested scenarios, it is acknowledged that the kind of signature it uses to recognize a library (i.e.~the list of printable strings in the file's data) is fragile in the sense that it can easily be obfuscated. This is however true for all techniques, to a lesser degree.

\newpage
\section{Acknowledgements}
I'd like to thank my supervisor at the Radboud University Nijmegen, dr.~ir.~Erik Poll, for giving me feedback along the way about this document and introducing me to Riscure in the first place. Additionally, I want to thank everyone at Riscure for being welcoming and for the feedback and conversations on this topic, and specifically my supervisors Martijn~Bogaard and Jeroen~Senden for their ideas and feedback that guided me in this project.

\newpage
\medskip
\bibliography{references.bib}

\begin{thebibliography}{10}

\bibitem{bloom1970space}
Burton~H Bloom.
\newblock Space/time trade-offs in hash coding with allowable errors.
\newblock {\em Communications of the ACM}, 13(7):422--426, 1970.

\bibitem{Cesare2011}
Silvio Cesare and Yang Xiang.
\newblock {Malware variant detection using similarity search over sets of
  control flow graphs}.
\newblock {\em Trust, Security and Privacy in Computing and Communications},
  pages 181--189, 2011.

\bibitem{conti2008visual}
Gregory Conti, Erik Dean, Matthew Sinda, and Benjamin Sangster.
\newblock Visual reverse engineering of binary and data files.
\newblock In {\em Visualization for Computer Security}, pages 1--17. Springer,
  2008.

\bibitem{dullien2005graph}
Thomas Dullien and Rolf Rolles.
\newblock Graph-based comparison of executable objects (english version).
\newblock {\em SSTIC}, 5:1--3, 2005.

\bibitem{Gheorghescu2005}
Marius Gheorghescu.
\newblock {An Automated Virus Classification System}.
\newblock {\em Virus Bulletin Conference}, (October):294--300, 2005.

\bibitem{IDAFLIRT}
{Hex-Rays}.
\newblock {IDA F.L.I.R.T. Technology: Overview}.
\newblock \url{https://www.hex-rays.com/products/ida/tech/flirt/index.shtml},
  2017.
\newblock Accessed: 2017-02-02.

\bibitem{Koret2016}
Joxean Koret.
\newblock {Cosa Nostra - A Graph Based Malware Clustering Toolkit}.
\newblock \url{https://github.com/joxeankoret/cosa-nostra}, 2016.
\newblock Accessed: 2016-12-07.

\bibitem{Kruegel2006}
Christopher Kruegel, Engin Kirda, Darren Mutz, William Robertson, and Giovanni
  Vigna.
\newblock {Polymorphic worm detection using structural information of
  executables}.
\newblock In {\em Lecture Notes in Computer Science (including subseries
  Lecture Notes in Artificial Intelligence and Lecture Notes in
  Bioinformatics)}, volume 3858 LNCS, pages 207--226, 2006.

\bibitem{Tian2009}
Ronghua Tian, Lynn Batten, Rafiqul Islam, and Steve Versteeg.
\newblock {An automated classification system based on the strings of trojan
  and virus families}.
\newblock In {\em 2009 4th International Conference on Malicious and Unwanted
  Software, MALWARE 2009}, pages 23--30. IEEE, 2009.

\bibitem{BinDiffZynamics}
{Zynamics}.
\newblock {BinDiff}.
\newblock \url{https://www.zynamics.com/bindiff.html}, 2016.
\newblock Accessed: 2016-12-21.

\end{thebibliography}

\newpage
\appendix
\section{Dataset}
This appendix shows the contents of the reference database used for the experiments in this paper. Library versions were compiled for one of three architectures: X86, MIPS or ARM. As can be seen below, not all versions of every library were obtained for every architecture. Unless stated otherwise, the set of reference library versions used for each experiment is a subset of this database, as obtained by filtering this list by the architecture of the sample file.

~\\

\scriptsize
\begin{itemize}[noitemsep,topsep=0pt,parsep=0pt,partopsep=0pt]
\item attr:
\begin{itemize}[noitemsep,topsep=0pt,parsep=0pt,partopsep=0pt]
  \item \textbf{MIPS}: 2.4.47-2
\end{itemize}

\item bzip2:
\begin{itemize}[noitemsep,topsep=0pt,parsep=0pt,partopsep=0pt]
  \item \textbf{ARM}: 1.0.4 1.0.5 1.0.6
  \item \textbf{MIPS}: 1.0.5-5 1.0.6-5
\end{itemize}

\item curl:
\begin{itemize}[noitemsep,topsep=0pt,parsep=0pt,partopsep=0pt]
  \item \textbf{ARM}: 7.21.4 7.21.5 7.21.6 7.21.7 7.22.0 7.23.0 7.23.1 7.24.0 7.25.0 7.26.0 7.27.0 7.28.0 7.28.1 7.29.0 7.30.0 7.31.0 7.32.0 7.33.0 7.34.0 7.35.0 7.36.0 7.37.0 7.37.1 7.38.0 7.39.0 7.40.0 7.41.0 7.42.1 7.43.0 7.44.0 7.45.0 7.46.0 7.47.0 7.47.1 7.48.0 7.49.0 7.49.1 7.50.0 7.50.1
  \item \textbf{MIPS}: 7.21.4 7.21.5 7.21.6 7.21.7 7.22.0 7.23.0 7.23.1 7.24.0 7.25.0 7.26.0 7.27.0 7.28.0 7.28.1 7.29.0 7.30.0 7.31.0 7.32.0 7.33.0 7.34.0 7.35.0 7.36.0 7.37.0 7.37.1 7.38.0 7.39.0 7.40.0 7.41.0 7.42.1 7.43.0 7.44.0 7.45.0 7.46.0 7.47.0 7.47.1 7.48.0 7.49.0 7.49.1 7.50.0 7.50.1 7.50.2 7.50.3
\end{itemize}

\item flac:
\begin{itemize}[noitemsep,topsep=0pt,parsep=0pt,partopsep=0pt]
  \item \textbf{MIPS}: 1.2.1 1.3.0 1.3.1
\end{itemize}

\item giflib:
\begin{itemize}[noitemsep,topsep=0pt,parsep=0pt,partopsep=0pt]
  \item \textbf{MIPS}: 5.0.4-2 5.0.5-1 5.0.6-1 5.1.0-1 5.1.1-1 5.1.2-1 5.1.4-1
\end{itemize}

\item id3lib:
\begin{itemize}[noitemsep,topsep=0pt,parsep=0pt,partopsep=0pt]
  \item \textbf{ARM}: 3.8.0
\end{itemize}

\item lame:
\begin{itemize}[noitemsep,topsep=0pt,parsep=0pt,partopsep=0pt]
  \item \textbf{MIPS}: 3.98.4 3.99.1 3.99.2 3.99.3 3.99.4 3.99.5 3.99
\end{itemize}

\item libao:
\begin{itemize}[noitemsep,topsep=0pt,parsep=0pt,partopsep=0pt]
  \item \textbf{ARM}: 0.8.8 1.0.0 1.1.0 1.2.0
\end{itemize}

\item libarchive:
\begin{itemize}[noitemsep,topsep=0pt,parsep=0pt,partopsep=0pt]
  \item \textbf{ARM}: 3.0.4 3.1.2 3.2.0 3.2.1 3.2.2
\end{itemize}

\item libart-lgpl:
\begin{itemize}[noitemsep,topsep=0pt,parsep=0pt,partopsep=0pt]
  \item \textbf{ARM}: 2.3.21
\end{itemize}

\item libburn:
\begin{itemize}[noitemsep,topsep=0pt,parsep=0pt,partopsep=0pt]
  \item \textbf{ARM}: 0.6.0.pl01 0.6.2.pl00 0.6.4.pl00 0.6.6.pl00 0.6.8.pl00 0.7.0.pl00 0.7.2.pl00 0.7.2.pl01 0.7.4.pl00 0.7.6.pl00 0.8.0.pl00 0.8.2.pl00 0.8.4.pl00 0.8.6.pl00 0.8.8.pl00 0.9.0.pl00 1.0.0.pl00 1.0.2.pl00 1.0.4.pl00 1.0.6.pl00 1.1.0.pl01 1.1.0 1.1.4 1.1.6 1.1.8 1.2.0 1.2.2 1.2.4 1.2.6 1.2.8 1.3.0.pl01 1.3.0 1.3.2 1.3.4 1.3.6.pl01 1.3.6 1.3.8 1.4.0 1.4.2.pl01 1.4.2 1.4.4 1.4.6
\end{itemize}

\item libcdaudio:
\begin{itemize}[noitemsep,topsep=0pt,parsep=0pt,partopsep=0pt]
  \item \textbf{ARM}: 0.99.12.p2 0.99.12
  \item \textbf{MIPS}: 0.99.12-4 0.99.12-7 0.99.12
\end{itemize}

\item libcddb:
\begin{itemize}[noitemsep,topsep=0pt,parsep=0pt,partopsep=0pt]
  \item \textbf{ARM}: 1.3.0 1.3.2
\end{itemize}

\item libcdio:
\begin{itemize}[noitemsep,topsep=0pt,parsep=0pt,partopsep=0pt]
  \item \textbf{ARM}: 0.82 0.83 0.90 0.92 0.93
\end{itemize}

\item libcrypto:
\begin{itemize}[noitemsep,topsep=0pt,parsep=0pt,partopsep=0pt]
  \item \textbf{X86}: 0.9.6 0.9.7 0.9.8
  \item \textbf{MIPS}: 1.0.0
\end{itemize}

\item libdatrie:
\begin{itemize}[noitemsep,topsep=0pt,parsep=0pt,partopsep=0pt]
  \item \textbf{ARM}: 0.1.2 0.1.4 0.2.10 0.2.1 0.2.2 0.2.3 0.2.4 0.2.5 0.2.6 0.2.8 0.2.9
\end{itemize}

\item libdc1394:
\begin{itemize}[noitemsep,topsep=0pt,parsep=0pt,partopsep=0pt]
  \item \textbf{ARM}: 2.1.0 2.1.2 2.1.3 2.2.1 2.2.3 2.2.4
\end{itemize}

\item libdca:
\begin{itemize}[noitemsep,topsep=0pt,parsep=0pt,partopsep=0pt]
  \item \textbf{ARM}: 0.0.5
\end{itemize}

\item libdiscid:
\begin{itemize}[noitemsep,topsep=0pt,parsep=0pt,partopsep=0pt]
  \item \textbf{ARM}: 0.3.0 0.3.2 0.4.1 0.5.0 0.5.1 0.5.2 0.6.1
\end{itemize}

\item libdmtx:
\begin{itemize}[noitemsep,topsep=0pt,parsep=0pt,partopsep=0pt]
  \item \textbf{ARM}: 0.7.4
\end{itemize}

\item libdv:
\begin{itemize}[noitemsep,topsep=0pt,parsep=0pt,partopsep=0pt]
  \item \textbf{ARM}: 1.0.0
\end{itemize}

\item libdvbpsi:
\begin{itemize}[noitemsep,topsep=0pt,parsep=0pt,partopsep=0pt]
  \item \textbf{ARM}: 0.1.6 0.1.7 0.2.0 0.2.1 0.2.2 1.1.0 1.1.1 1.1.2 1.3.0
\end{itemize}

\item libdvdcss:
\begin{itemize}[noitemsep,topsep=0pt,parsep=0pt,partopsep=0pt]
  \item \textbf{ARM}: 1.2.10 1.2.11 1.2.12 1.2.13 1.2.9 1.3.0 1.4.0
\end{itemize}

\item libebml:
\begin{itemize}[noitemsep,topsep=0pt,parsep=0pt,partopsep=0pt]
  \item \textbf{ARM}: 0.7.8 0.8.0 1.0.0 1.2.0 1.2.1 1.2.2 1.3.0 1.3.1 1.3.3 1.3.4
  \item \textbf{MIPS}: 0.7.8-2 0.8.0-1 1.0.0-1 1.2.0-1 1.2.1-1 1.2.2-2 1.3.0-1 1.3.1-1 1.3.3-2 1.3.4-1 1.3.4
\end{itemize}

\item libexif:
\begin{itemize}[noitemsep,topsep=0pt,parsep=0pt,partopsep=0pt]
  \item \textbf{ARM}: 0.6.16 0.6.17 0.6.19 0.6.20 0.6.21
  \item \textbf{MIPS}: 0.6.17-1 0.6.19-1 0.6.20-1 0.6.21-3
\end{itemize}

\item libffi:
\begin{itemize}[noitemsep,topsep=0pt,parsep=0pt,partopsep=0pt]
  \item \textbf{ARM}: 3.0.10 3.0.11 3.0.12 3.0.13 3.0.8 3.0.9 3.1 3.2.1
\end{itemize}

\item libgssglue:
\begin{itemize}[noitemsep,topsep=0pt,parsep=0pt,partopsep=0pt]
  \item \textbf{ARM}: 0.1 0.3 0.4
\end{itemize}

\item libid3tag:
\begin{itemize}[noitemsep,topsep=0pt,parsep=0pt,partopsep=0pt]
  \item \textbf{ARM}: 0.15.1b
  \item \textbf{MIPS}: 0.15.1b
\end{itemize}

\item libieee1284:
\begin{itemize}[noitemsep,topsep=0pt,parsep=0pt,partopsep=0pt]
  \item \textbf{ARM}: 0.2.11
\end{itemize}

\item libisoburn:
\begin{itemize}[noitemsep,topsep=0pt,parsep=0pt,partopsep=0pt]
  \item \textbf{ARM}: 1.1.6 1.1.8 1.2.0 1.2.2 1.2.4 1.2.6 1.2.8 1.3.0 1.3.2 1.3.4 1.3.6.pl01 1.3.6 1.3.8 1.4.0 1.4.2 1.4.4 1.4.6
\end{itemize}

\item libisofs:
\begin{itemize}[noitemsep,topsep=0pt,parsep=0pt,partopsep=0pt]
  \item \textbf{ARM}: 0.6.12 0.6.14 0.6.16 0.6.18 0.6.20 0.6.22 0.6.24 0.6.26 0.6.28 0.6.30 0.6.32 0.6.34 0.6.36 0.6.38 0.6.40 1.0.0 1.0.2 1.0.4 1.0.6 1.0.8 1.1.0 1.1.2 1.1.4 1.1.6 1.2.0 1.2.2 1.2.4 1.2.6 1.2.8 1.3.0 1.3.2 1.3.4 1.3.6 1.3.8 1.4.0 1.4.2 1.4.4 1.4.6
\end{itemize}

\item libjpeg:
\begin{itemize}[noitemsep,topsep=0pt,parsep=0pt,partopsep=0pt]
  \item \textbf{MIPS}: 7.0.0 8.0.0 8.0.1 8.0.2 8.3.0 8.4.0 9.0.0 9.1.0 9.2.0
  \item \textbf{X86}: 7.0.0 8.0.0 8.0.1 8.0.2 8.3.0 8.4.0 9.0.0 9.1.0 9.2.0
\end{itemize}

\item liblo:
\begin{itemize}[noitemsep,topsep=0pt,parsep=0pt,partopsep=0pt]
  \item \textbf{ARM}: 0.23 0.25 0.26 0.27 0.28
\end{itemize}

\item libmcrypt:
\begin{itemize}[noitemsep,topsep=0pt,parsep=0pt,partopsep=0pt]
  \item \textbf{ARM}: 2.5.8
  \item \textbf{MIPS}: 2.5.8-2 2.5.8
\end{itemize}

\item libmms:
\begin{itemize}[noitemsep,topsep=0pt,parsep=0pt,partopsep=0pt]
  \item \textbf{ARM}: 0.6.4
\end{itemize}

\item libmodplug:
\begin{itemize}[noitemsep,topsep=0pt,parsep=0pt,partopsep=0pt]
  \item \textbf{ARM}: 0.8.7 0.8.8.1 0.8.8.2 0.8.8.3 0.8.8.4 0.8.8.5 0.8.8
\end{itemize}

\item libmp4v2:
\begin{itemize}[noitemsep,topsep=0pt,parsep=0pt,partopsep=0pt]
  \item \textbf{ARM}: 2.0.0
\end{itemize}

\item libmpdclient:
\begin{itemize}[noitemsep,topsep=0pt,parsep=0pt,partopsep=0pt]
  \item \textbf{ARM}: 2.10 2.8 2.9
\end{itemize}

\item libmpeg2:
\begin{itemize}[noitemsep,topsep=0pt,parsep=0pt,partopsep=0pt]
  \item \textbf{ARM}: 0.4.1 0.5.1
\end{itemize}

\item libmspack:
\begin{itemize}[noitemsep,topsep=0pt,parsep=0pt,partopsep=0pt]
  \item \textbf{ARM}: 0.0.20060920alpha 0.2alpha 0.3alpha 0.4alpha 0.5alpha
\end{itemize}

\item libnet:
\begin{itemize}[noitemsep,topsep=0pt,parsep=0pt,partopsep=0pt]
  \item \textbf{ARM}: 1.1.6
\end{itemize}

\item libnova:
\begin{itemize}[noitemsep,topsep=0pt,parsep=0pt,partopsep=0pt]
  \item \textbf{ARM}: 0.12.1 0.12.2 0.12.3 0.13.0 0.14.0 0.15.0
\end{itemize}

\item libogg:
\begin{itemize}[noitemsep,topsep=0pt,parsep=0pt,partopsep=0pt]
  \item \textbf{ARM}: 1.1.3 1.1.4 1.2.0 1.2.1 1.2.2 1.3.0 1.3.1 1.3.2
  \item \textbf{MIPS}: 1.1.3 1.1.4 1.2.0 1.2.1 1.2.2 1.3.0 1.3.1 1.3.2
\end{itemize}

\item liboil:
\begin{itemize}[noitemsep,topsep=0pt,parsep=0pt,partopsep=0pt]
  \item \textbf{ARM}: 0.3.17
\end{itemize}

\item libpciaccess:
\begin{itemize}[noitemsep,topsep=0pt,parsep=0pt,partopsep=0pt]
  \item \textbf{ARM}: 0.10.4 0.10.5 0.10.6 0.10.8 0.10.9 0.11.0 0.12.0 0.12.1 0.12.902 0.13.1 0.13.2 0.13.3 0.13.4 0.13
\end{itemize}

\item libpipeline:
\begin{itemize}[noitemsep,topsep=0pt,parsep=0pt,partopsep=0pt]
  \item \textbf{ARM}: 1.2.2 1.2.3 1.2.4 1.2.5 1.2.6 1.3.0 1.3.1 1.4.0 1.4.1
\end{itemize}

\item libpng:
\begin{itemize}[noitemsep,topsep=0pt,parsep=0pt,partopsep=0pt]
  \item \textbf{X86}: 2.1.0.12 3.1.2.12 3.1.2.18 3.1.2.1 3.1.2.27 3.1.2.32 3.1.2.37 3.1.2.5 3.1.2.8 3.43.0 3.44.0 3.50.0
  \item \textbf{ARM}: 1.2.25 1.2.29 1.2.30 1.2.31 1.2.32 1.4.3 1.4.4 1.4.5 1.4.8 1.5.10 1.5.11 1.5.12 1.5.13 1.5.14 1.5.15 1.5.1 1.5.7 1.5.8 1.5.9 1.6.10 1.6.12 1.6.13 1.6.14 1.6.15 1.6.16 1.6.18 1.6.19 1.6.20 1.6.21 1.6.22 1.6.23 1.6.24 1.6.25 1.6.26 1.6.2 1.6.3 1.6.5 1.6.6 1.6.7 1.6.8 1.6.9
  \item \textbf{MIPS}: 1.5.10 1.5.11 1.5.12 1.5.13 1.5.14 1.5.15 1.6.10 1.6.12 1.6.13 1.6.14 1.6.15 1.6.16 1.6.18 1.6.19 1.6.20 1.6.21 1.6.22 1.6.23 1.6.24 1.6.25 1.6.26 1.6.2 1.6.3 1.6.5 1.6.6 1.6.7 1.6.8 1.6.9
\end{itemize}

\item libraw1394:
\begin{itemize}[noitemsep,topsep=0pt,parsep=0pt,partopsep=0pt]
  \item \textbf{ARM}: 2.0.4 2.0.5 2.0.7 2.1.0 2.1.1 2.1.2
  \item \textbf{MIPS}: 2.0.4 2.0.5 2.0.7 2.1.0 2.1.1 2.1.2
\end{itemize}

\item libsamplerate:
\begin{itemize}[noitemsep,topsep=0pt,parsep=0pt,partopsep=0pt]
  \item \textbf{ARM}: 0.1.2 0.1.3 0.1.4 0.1.6 0.1.7 0.1.8 0.1.9
\end{itemize}

\item libsigsegv:
\begin{itemize}[noitemsep,topsep=0pt,parsep=0pt,partopsep=0pt]
  \item \textbf{ARM}: 2.10 2.4 2.6
\end{itemize}

\item libsndfile:
\begin{itemize}[noitemsep,topsep=0pt,parsep=0pt,partopsep=0pt]
  \item \textbf{ARM}: 1.0.21 1.0.22 1.0.23 1.0.24 1.0.25 1.0.26 1.0.27
\end{itemize}

\item libssl:
\begin{itemize}[noitemsep,topsep=0pt,parsep=0pt,partopsep=0pt]
  \item \textbf{MIPS}: 1.0.0
  \item \textbf{X86}: 0.9.6 0.9.7 0.9.8 1.0.0
\end{itemize}

\item libtar:
\begin{itemize}[noitemsep,topsep=0pt,parsep=0pt,partopsep=0pt]
  \item \textbf{MIPS}: 1.2.20
\end{itemize}

\item libtiff:
\begin{itemize}[noitemsep,topsep=0pt,parsep=0pt,partopsep=0pt]
  \item \textbf{ARM}: 4.0.7
  \item \textbf{MIPS}: 3.8.2 3.9.0 3.9.1 3.9.2 3.9.4 3.9.5 4.0.0 4.0.1 4.0.2 4.0.3 4.0.4 4.0.6 4.0.7
\end{itemize}

\item libupnp:
\begin{itemize}[noitemsep,topsep=0pt,parsep=0pt,partopsep=0pt]
  \item \textbf{ARM}: 1.6.10 1.6.12 1.6.13 1.6.14 1.6.15 1.6.16 1.6.17 1.6.18 1.6.19 1.6.20 1.6.6 1.6.8 1.6.9
\end{itemize}

\item libvisual:
\begin{itemize}[noitemsep,topsep=0pt,parsep=0pt,partopsep=0pt]
  \item \textbf{ARM}: 0.4.0
\end{itemize}

\item libvncserver:
\begin{itemize}[noitemsep,topsep=0pt,parsep=0pt,partopsep=0pt]
  \item \textbf{ARM}: 0.9.10 0.9.1 0.9.7 0.9.8.1 0.9.8.2 0.9.8 0.9.9
\end{itemize}

\item libxmi:
\begin{itemize}[noitemsep,topsep=0pt,parsep=0pt,partopsep=0pt]
  \item \textbf{ARM}: 1.2
\end{itemize}

\item libytnef:
\begin{itemize}[noitemsep,topsep=0pt,parsep=0pt,partopsep=0pt]
  \item \textbf{ARM}: 1.5 1.8
\end{itemize}

\item libz:
\begin{itemize}[noitemsep,topsep=0pt,parsep=0pt,partopsep=0pt]
  \item \textbf{ARM}: 1.2.8
  \item \textbf{MIPS}: 1.0.4 1.1.2 1.1.3 1.2.2 1.2.3 1.2.8
  \item \textbf{X86}: 1.0.8 1.0.9 1.1.0 1.1.1 1.1.2 1.1.3 1.1.4 1.2.0.1 1.2.0.2 1.2.0.3 1.2.0.4 1.2.0.5 1.2.0.6 1.2.0.7 1.2.0.8 1.2.0 1.2.1.1 1.2.1.2 1.2.1 1.2.2.1 1.2.2.2 1.2.2.3 1.2.2.4 1.2.2 1.2.3.1 1.2.3.2 1.2.3.3 1.2.3.4 1.2.3.5 1.2.3.6 1.2.3.7 1.2.3.8 1.2.3.9 1.2.3 1.2.4.1 1.2.4.2 1.2.4.3 1.2.4.4 1.2.4.5 1.2.4 1.2.5.2 1.2.5.3 1.2.5 1.2.6.1 1.2.6 1.2.7.1 1.2.7.2-motley 1.2.7.3 1.2.7
\end{itemize}

\item libzip:
\begin{itemize}[noitemsep,topsep=0pt,parsep=0pt,partopsep=0pt]
  \item \textbf{ARM}: 0.10.1 0.10 0.11.1 0.11.2 0.11 0.8 0.9.3 0.9 1.0.1 1.1.2 1.1.3
\end{itemize}

\item lzo:
\begin{itemize}[noitemsep,topsep=0pt,parsep=0pt,partopsep=0pt]
  \item \textbf{ARM}: 1.08 2.08 2.09
  \item \textbf{MIPS}: 1.08-5 1.08-8 2.08-3 2.09-1
\end{itemize}

\item opus:
\begin{itemize}[noitemsep,topsep=0pt,parsep=0pt,partopsep=0pt]
  \item \textbf{ARM}: 1.0.1 1.0.2 1.0.3 1.1.1 1.1.2 1.1.3 1.1
  \item \textbf{MIPS}: 1.0.1 1.0.2 1.0.3 1.1.1 1.1.2 1.1.3 1.1
\end{itemize}

\item pcre:
\begin{itemize}[noitemsep,topsep=0pt,parsep=0pt,partopsep=0pt]
  \item \textbf{ARM}: 8.39
\end{itemize}

\item uClibc:
\begin{itemize}[noitemsep,topsep=0pt,parsep=0pt,partopsep=0pt]
  \item \textbf{MIPS}: 0.9.27 0.9.28 0.9.29 0.9.30.1 0.9.33.2
\end{itemize}

\item xz:
\begin{itemize}[noitemsep,topsep=0pt,parsep=0pt,partopsep=0pt]
  \item \textbf{ARM}: 5.0.0 5.0.1 5.0.2 5.0.3 5.0.4 5.0.5 5.0.6 5.0.7 5.0.8 5.2.0 5.2.1 5.2.2
  \item \textbf{MIPS}: 5.0.0 5.0.1 5.0.2 5.0.3 5.0.4 5.0.5 5.0.6 5.0.7 5.0.8 5.2.0 5.2.1 5.2.2
\end{itemize}
\end{itemize}

\end{document}